\documentclass{emulateapj}
\shortauthors{Wang et al.}

\newcommand{\spz}{\textit{Spitzer}}

\begin{document}

\title{Searching for Debris Disks around Seven Radio Pulsars}

\author{Zhongxiang Wang\altaffilmark{1}, 
C.-Y. Ng\altaffilmark{2},
Xuebing Wang\altaffilmark{1,3},
Aigen Li\altaffilmark{4},
David L. Kaplan\altaffilmark{5}
}

\altaffiltext{1}{\footnotesize 
Shanghai Astronomical Observatory, Chinese Academy of Sciences,
80 Nandan Road, Shanghai 200030, China}

\altaffiltext{2}{Department of Physics, The University of Hong Kong,
Pokfulam Road, Hong Kong}

\altaffiltext{3}{Graduate University of Chinese Academy of Sciences, 
Beijing 100049, China}

\altaffiltext{4}{Department of Physics and Astronomy, University of Missouri, 
Columbia, MO 65211, USA}

\altaffiltext{5}{Physics Department, University of Wisconsin—Milwaukee, 
Milwaukee, WI 53211, USA}

\begin{abstract}
We report on our searches for debris disks around seven relatively nearby 
radio pulsars, which are isolated sources and were carefully selected as 
the targets on the basis of our deep $K_s$-band imaging survey. 
The $K_s$ images obtained with the 6.5\,m Baade Magellan Telescope 
at Las Campanas Observatory are analyzed together with 
the \textit{Spitzer}/IRAC images at 4.5 and 8.0~$\mu$m
and the \textit{WISE} images at 3.4, 4.6, 12 and 22~$\mu$m.
No infrared (IR) counterparts of these pulsars are found, 
with flux upper limits of $\sim \mu$Jy at near-infrared ($\lambda<10\ \mu$m) 
and $\sim$10--1000\,$\mu$Jy 
at mid-infrared wavelengths ($\lambda>10\ \mu$m). 
The results of this search are discussed in terms of 
the efficiency of converting the pulsar spin-down energy to thermal energy
and X-ray heating of debris disks, with comparison made to
the two magnetars 4U~0142+61 and 
1E~2259+586 which are suggested to harbor a debris disk.

\end{abstract}


\section{INTRODUCTION}

Disks are ubiquitous in astrophysical systems at all scales, found
almost anywhere an axis of symmetry exists, and play an important role
in the appearance and evolution of these systems. For example, they
are essential for the creation of planetary systems, formation of
young stars, mass-transfer in binary systems, and fueling of the
central engines in active galactic nuclei, even though the physical
properties of both the disks and their host systems differ dramatically.

Disks may also exist around isolated neutron stars, possibly 
formed from fallback
material after supernova explosions \citep{lwb91,chn00}. 
When a massive star dies in a supernova explosion, 
some amount of the ejecta ($\leq 0.1 M_{\odot}$) may ``fallback'' onto 
the newly formed neutron star \citep{che89,ww95}.
If much of the material is captured by the gravitational field
of the neutron star and has sufficiently high angular momentum to prevent
direct infall onto this compact star, a disk forms 
due to interactions between streams of the captured material.

There are a number of observational puzzles relating to radio pulsars 
that could be explained by disks \citep{md81,cs08}.  The evolution of
pulse periods with time for most pulsars does not actually follow the
detailed predictions of magnetic dipole radiation 
(e.g., \citealt{liv+07,esp+11}), which could be due to disks 
\citep{mph01,cal+13} exerting excess torque on the pulsars' magnetospheres.  
In addition, some unique phenomena seen in individual pulsars, such as jets
and neutron star precession, may arise from disks \citep{bp04,qia+03}.
The first extra-solar planet system was
discovered around the radio pulsar B1257+12 \citep{wf92}. 
Even though the source is a millisecond pulsar with a
complicated evolutionary path, it is generally believed that the
planetary system had formed from a debris disk (e.g., \citealt{ph93,mh01}).  
There is a type of radio emitting neutron stars,
called rotating radio transients (RRATs; \citealt{mcl+06}),
which are characterized by short, periodic radio bursts (2--30 ms) 
spaced by long intervals (ranging from minutes to hours).
The bursts, along with other
variations in radio pulses---such as pulse nulling, drifting subpulses,
and mode changes, might be caused by dusty material from a
surrounding debris disk falling into the magnetosphere of a pulsar
\citep{cs08,sha+13}.

Motivated by the discovery of the planetary system around PSR B1257+12, 
searches for debris disks around neutron stars have concentrated on
old radio pulsars, but with no 
success (e.g. \citealt{ff96,koc+02,lf04}).
These searches were however limited by the
choice of targets (older, less energetic pulsars) and the
instrumentation (poor sensitivity and angular resolution).  
Multi-wavelength searches for fallback disks around
a number of young neutron stars have been conducted 
(e.g., \citealt{wck06,wkc07}), although these 
are considered as high-energy objects,
namely, central compact objects (CCOs; \citealt{pst04,gha13}) and 
magnetars \citep{td96}, rather than ordinary/traditional radio pulsars.
The former are a class of radio-quiet X-ray point
sources at the centers of young supernova remnants 
while the latter are believed to generally have
ultra-high surface magnetic fields ($\sim10^{14}$~G; 
\citealt{wt06}).

Using the {\it Spitzer Space Telescope}, the infrared (IR) 
counterparts to the
magnetars 4U 0142+61 and 1E~2259+586 have been found \citep{wck06,kap+09}. 
For the former source, no single spectral model, such as a blackbody or 
a power law, can
fit its combined optical and IR spectral energy distribution (SED).
Instead, it is very likely that the IR emission arises from a
residual dust disk.  This disk, with estimated mass of $<$10$^{-3}
M_{\odot}$, is heated by strong X-rays from the central pulsar
($L_{\rm X}\simeq 10^{36}$ erg s$^{-1}$) and appears bright at IR
wavelengths (for example, $L_{\rm 4.5\mu m}\simeq 5\times 10^{31}$
ergs s$^{-1}$).  The disk's lifetime was estimated to be $>10^6$~yr, assuming
that the magnetar has been spun down by mass accretion from the putative disk
\citep{wck06}.  The lifetime significantly
exceeds the pulsar age ($\approx 10^5$ yr), suggesting a
supernova fallback origin \citep{wck06}. While there were 
only $K_s$-band and \spz\ 4.5 $\mu$m
detections of 1E~2259+586, the overall IR emission is similar to that
of 4U 0142+61, suggesting the possible existence of a debris 
disk around this magnetar \citep{kap+09}. 

%%Recently the Vela pulsar was also found to have
%%similar emission excess at \spz\ 3.6 and 5.8 $\mu$m, and the possible 
%%indication of an X-ray irradiated disk around the pulsar was discussed
%%\citep{dan+11}.
%%Recently \citet{dan+11} have found a candidate IR counterpart to the Vela pulsar from \spz\ imaging at 3.6 and 5.8~$\mu$m. If it is confirmed, the IR emission would likely indicate the existence of a debris disk around the middle-aged (spin-down age is 11 kyr) pulsar and provide a certain case for our understanding of the appearance of debris disks around canonical radio pulsars. The Vela pulsar has $\dot{E_{\rm sd}}=6.9\times 10^{36}$ erg s$^{-1}$ and $L_{\rm X}\simeq 5.3\times 10^{32}$ \citep{pav+01} and is at a distance of 287~pc \citep{dod+03}, which imply fractions of 2.4$\times 10^{-8}$ or 3.2$\times 10^{-4}$ (at 5.8~$\mu$m) when its $\dot{E}_{\rm sd}$ or X-ray emission is considered, respectively. The value from considering X-ray heating is very similar to that in the 4U~0142+61 case, suggesting that X-ray heating probably plays the key role in making a debris disk detectable. Therefore the non-detection of any disks in this search could be just due to insufficient X-ray heating.

In order to explore the general existence of similar residual disks around 
isolated neutron stars, we have carried out disk searches around a few
relatively young radio pulsars.  The detectability of the disk
around 4U~0142+61 is likely the result of a combination of the very
strong X-ray heating of the disk and the relative youth of the object.
For radio pulsars, their winds that carry much of their spin-down
luminosity and contain energetic particles 
may heat a debris disk, producing IR emission \citep{jon07,jon08}.
Some fraction of the spin-down luminosity is conceivably assumed to illuminate
surrounding disks.

In this paper we report the results of disk searches around seven 
relatively young pulsars from our ground-based and \spz\ observations. 
The pulsar targets were carefully selected partly on
the basis of an initial near-IR $K_s$-band survey. 
Results from the all-sky survey conducted by
the \textit{Wide-field Infrared Survey Explorer} (\textit{WISE}) were included.
In addition, in order to understand the role of X-ray emission from neutron
stars in disk heating (or disk detectability), X-ray flux
values or upper limits of the seven pulsars, either from the literature 
or archival data analysis conducted by us, were also reported.  Below we first 
describe our target selection
in Section~\ref{subsec:ts}. The observations and data analysis are
described in Section~\ref{sec:obs}, and the results
and discussion are presented in Sections~\ref{sec:res} and \ref{sec:dis},
respectively.

\subsection{Target Selection}
\label{subsec:ts}
We selected sources from the Pulsar Catalog\footnote{http://www.atnf.csiro.au/people/pulsar/psrcat/} \citep{man+05}
based on the spin-down luminosity $L_{\rm sd}$  
scaled by distance $d$, $L_{\rm sd}/d^2$, and picked the 
top 40 young, energetic pulsars that had not been well observed in the 
optical/IR bands.  As one might expect near-IR emission from 
these systems if debris disks are present, we undertook observations 
in the $K_s$ band
using the 6.5-m Magellan telescopes (see Section~\ref{sec:obs}).
However, in no case did we find a near-IR counterpart.

Based on the ground-based observations, we then selected seven pulsars 
(see Table~\ref{tab:tgt}) as 
our targets for \spz\ observations. The reasons for the selection are the
following.  First, they
are relatively nearby, with a typical distance of 4 kpc.  A debris disk 
with temperature of approximately 1000~K \citep{wck06} would be bright 
in the mid-IR at that distance.  
Second, these targets are not surrounded by synchrotron nebulae that
can envelope some of the most energetic pulsars \citep{krh06,gs06} 
and could hide mid-IR emission from the pulsar itself.
Third and most important, since most pulsars are located primarily in
the Galactic plane, the confusion with field stars is a serious
problem.  These targets were selected because they are in less crowded
fields.  Using the Galactic disk
dust map \citep{sfd98}, we estimated the extinction along each 
source's direction. The maximum extinction for our targets is approximately
$A_V=9$ (for three high Galactic latitude sources, they only have
$A_V\approx 2$). This limits the extinction effect that could alter
the apparent stellar colors and cause us to mistake it for a disk.

To our target selection we added one other source: RRAT J1317$-$5759.
This object, which just emits sporadic bursts of radio waves,
would not be detected simply based on our spin-down luminosity
argument. However, the mechanism for the radio bursting is not clearly
understood, and may be related to a surrounding debris disk \citep{cs08}. 
Given that this source also has a clean field, we
selected it as our target for testing this hypothesis.

\section{OBSERVATIONS AND DATA REDUCTION}    % Section 2 
\label{sec:obs}

\subsection{Magellan $K_s$-band Imaging}

We observed the fields of the seven pulsar targets in the $K_s$ band
using Persson's Auxiliary Nasmyth Infrared Camera (PANIC; \citealt{mpm+04})
on the 6.5-m Baade Magellan Telescope at Las Campanas Observatory in Chile.
The observation dates are provided in Table~\ref{tab:obs}.
The detector was a Rockwell Hawaii 1024$\times$1024 HgCdTe array, having
a field of view (FOV) of 2\arcmin$\times$2\arcmin\  
and a pixel scale of 0.125\arcsec\ pixel$^{-1}$. The total on-source exposure 
times were in a range of 13--30 min. During the exposure of a target, 
the telescope was dithered in a
3$\times$3 grid with offsets of 10\arcsec\ to obtain a measurement of
the sky background. The observing conditions were good, 
with 0\farcs4--0\farcs5 seeing in the $K_s$ band. The FWHMs of point 
sources in our images are given in Table~\ref{tab:obs}.

We used the IRAF data analysis package for data reduction.
The images were bias-subtracted and flat-fielded. From each set of 
dithered images in one observation, 
a sky image was made by filtering out stars. The sky image was
subtracted from the set of images, and then the sky-subtracted images
were shifted and combined into one final image of a target field.

We astrometrically calibrated the images using the Two Micron
All Sky Survey (2MASS; \citealt{2mass}) stars detected in the fields.
For the targets except B0950+08, 
20--60 2MASS stars were used for the calibration, and the resulting  
uncertainties are of the order of 0\farcs03. Therefore, the
positional uncertainties for locating the six pulsars on the images are 
dominated by the 2MASS systematic uncertainty of $\simeq$0\farcs15.
For B0950+08, only six 2MASS stars were detected to be used for 
the calibration, because it has a high Galactic latitude $Gb$.
The nominal position uncertainty is 0\farcs13. Adding the 2MASS systematic
uncertainty in quadrature, the positional uncertainty for locating it
on the $K_s$ image is 0\farcs2.
\begin{figure*}
\centering
\begin{tabular}{c c}
\includegraphics[scale=0.35]{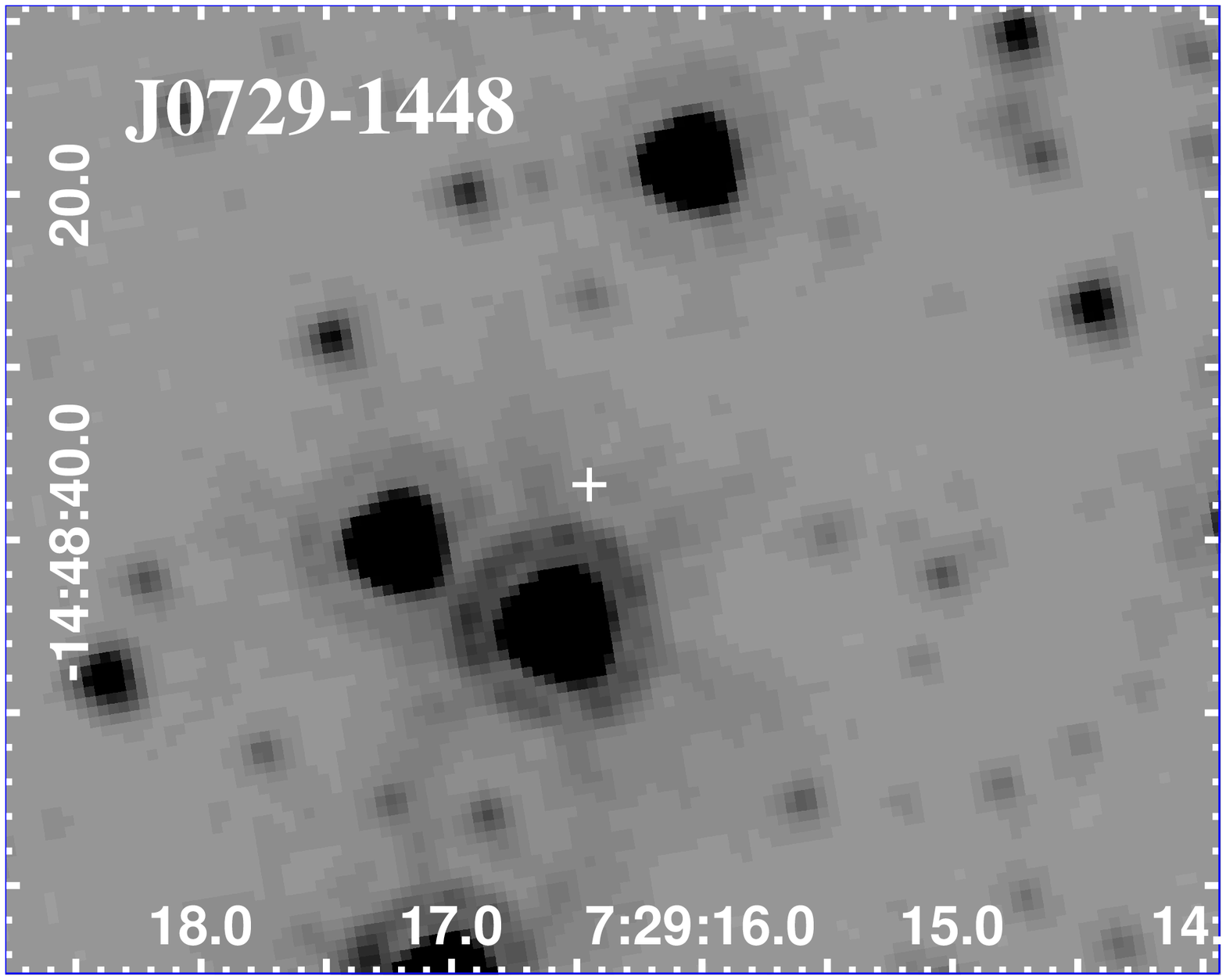}
& \includegraphics[scale=0.35]{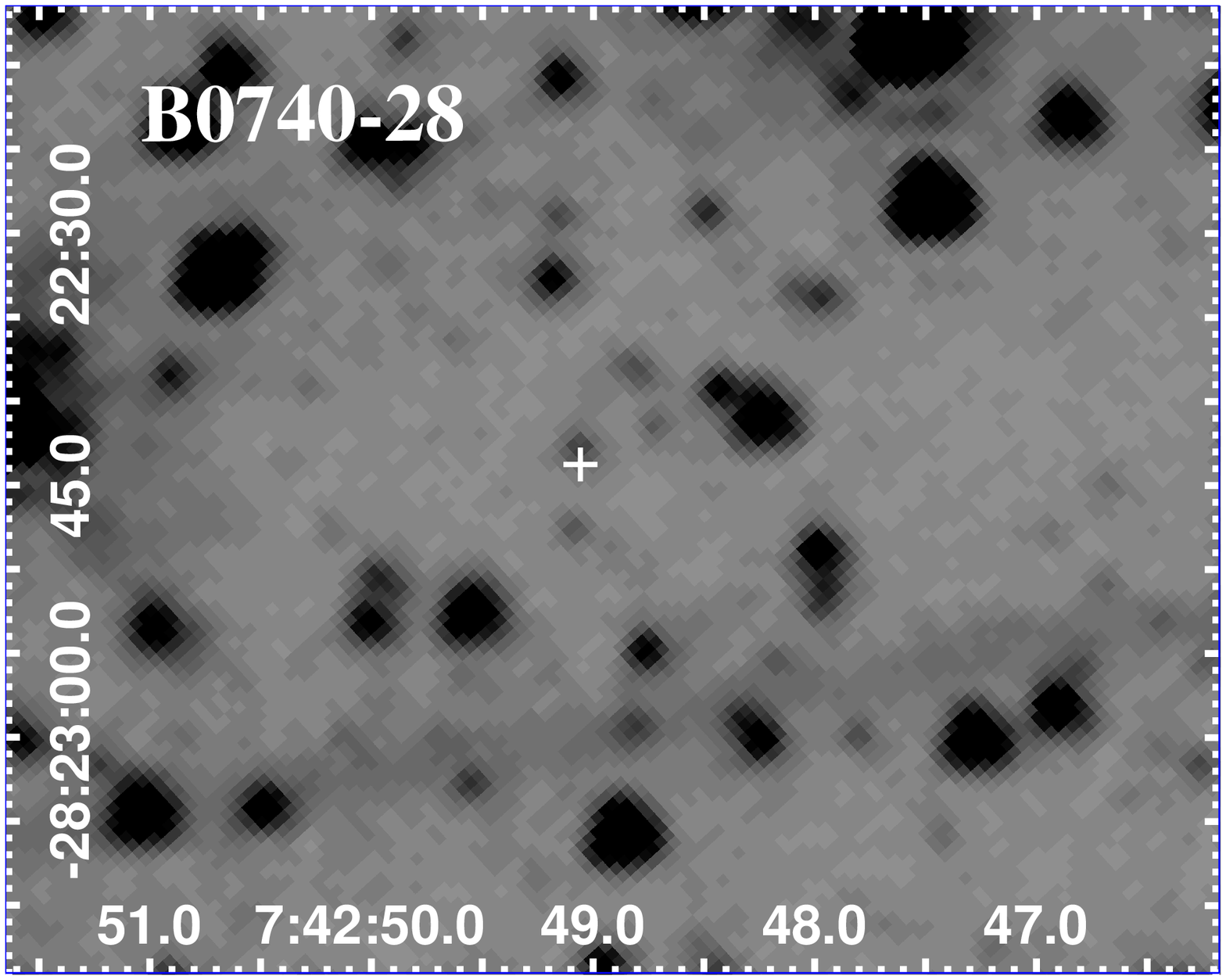} \\
\includegraphics[scale=0.35]{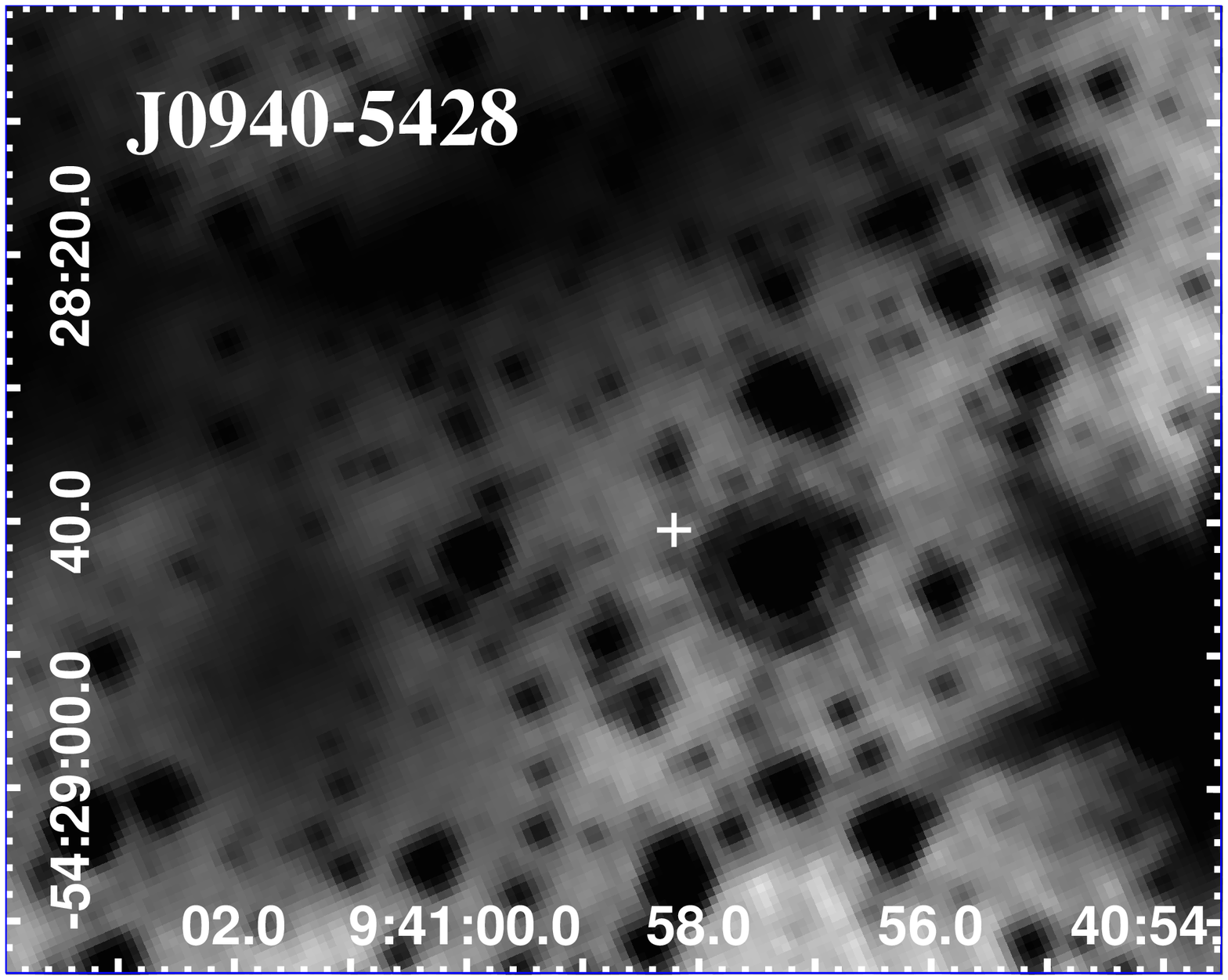} & 
\includegraphics[scale=0.35]{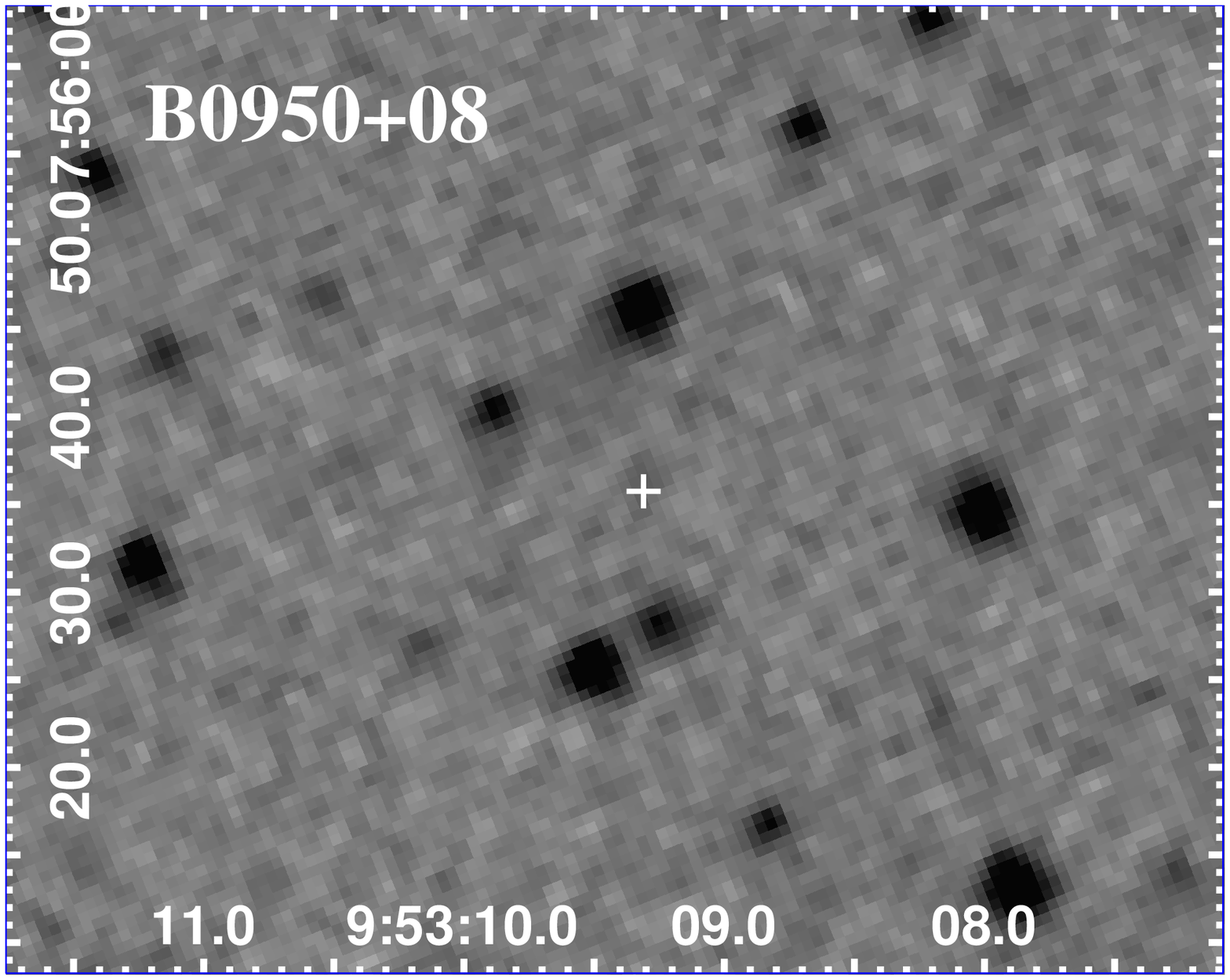} \\
\includegraphics[scale=0.35]{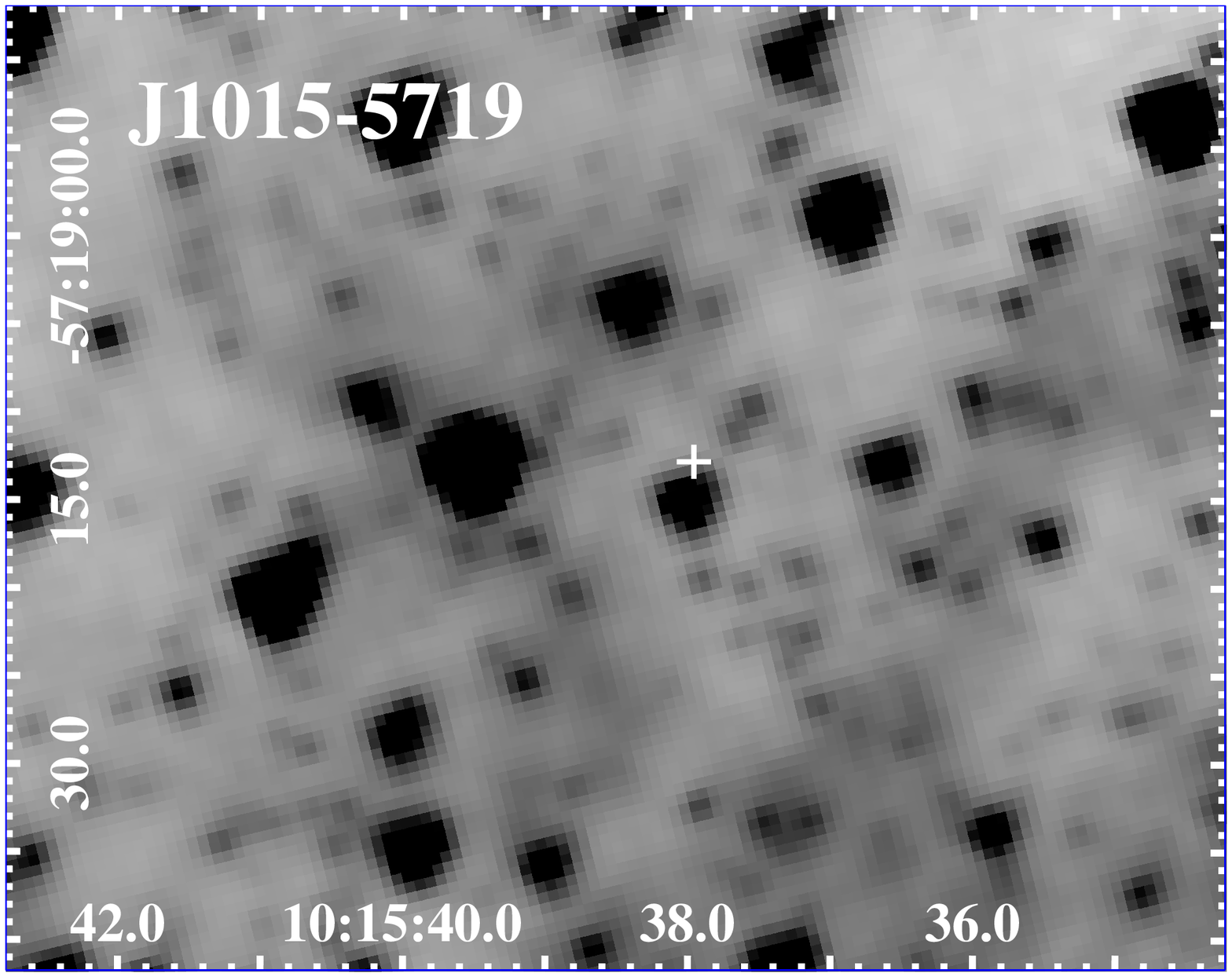} & 
\includegraphics[scale=0.35]{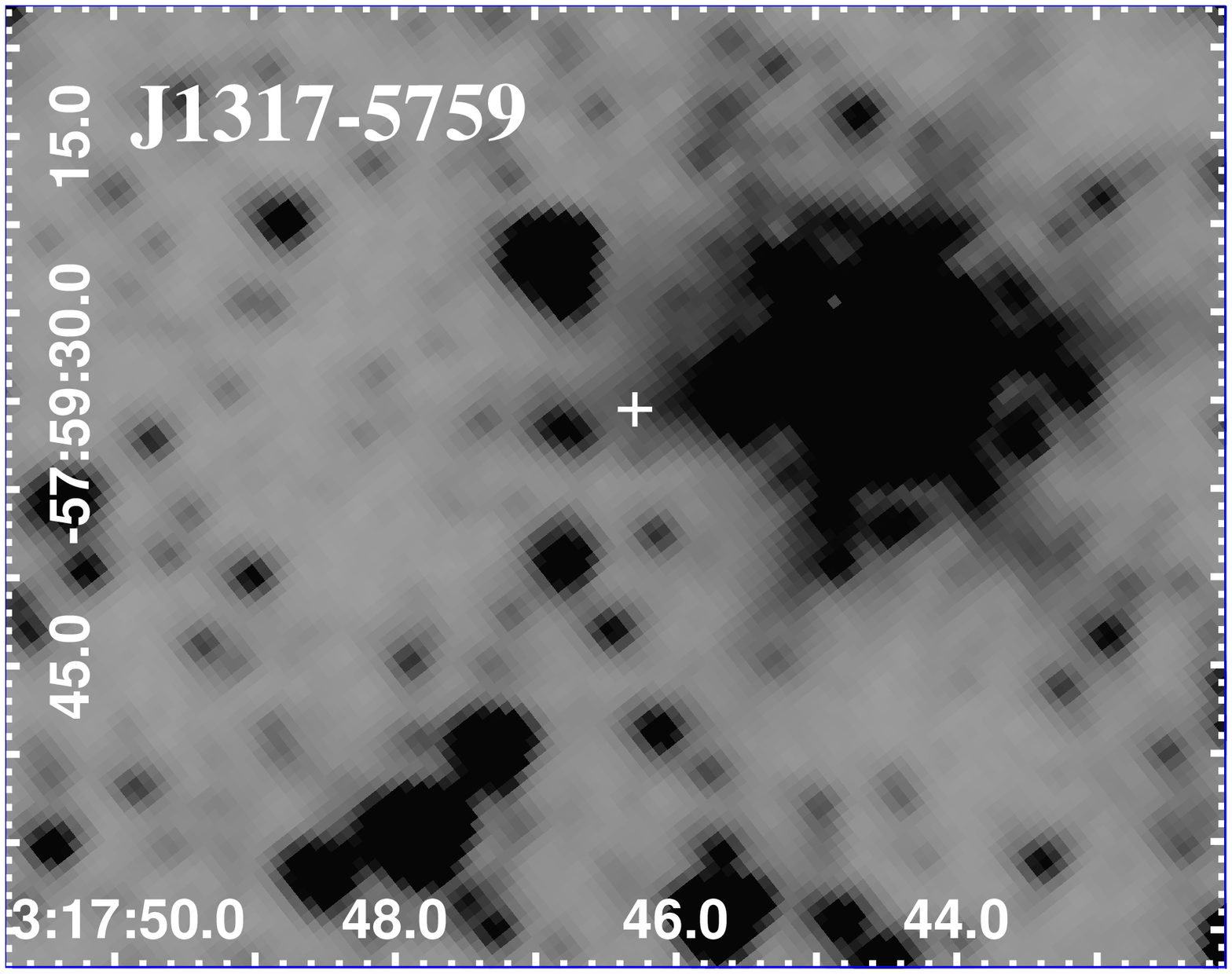} \\
\includegraphics[scale=0.35]{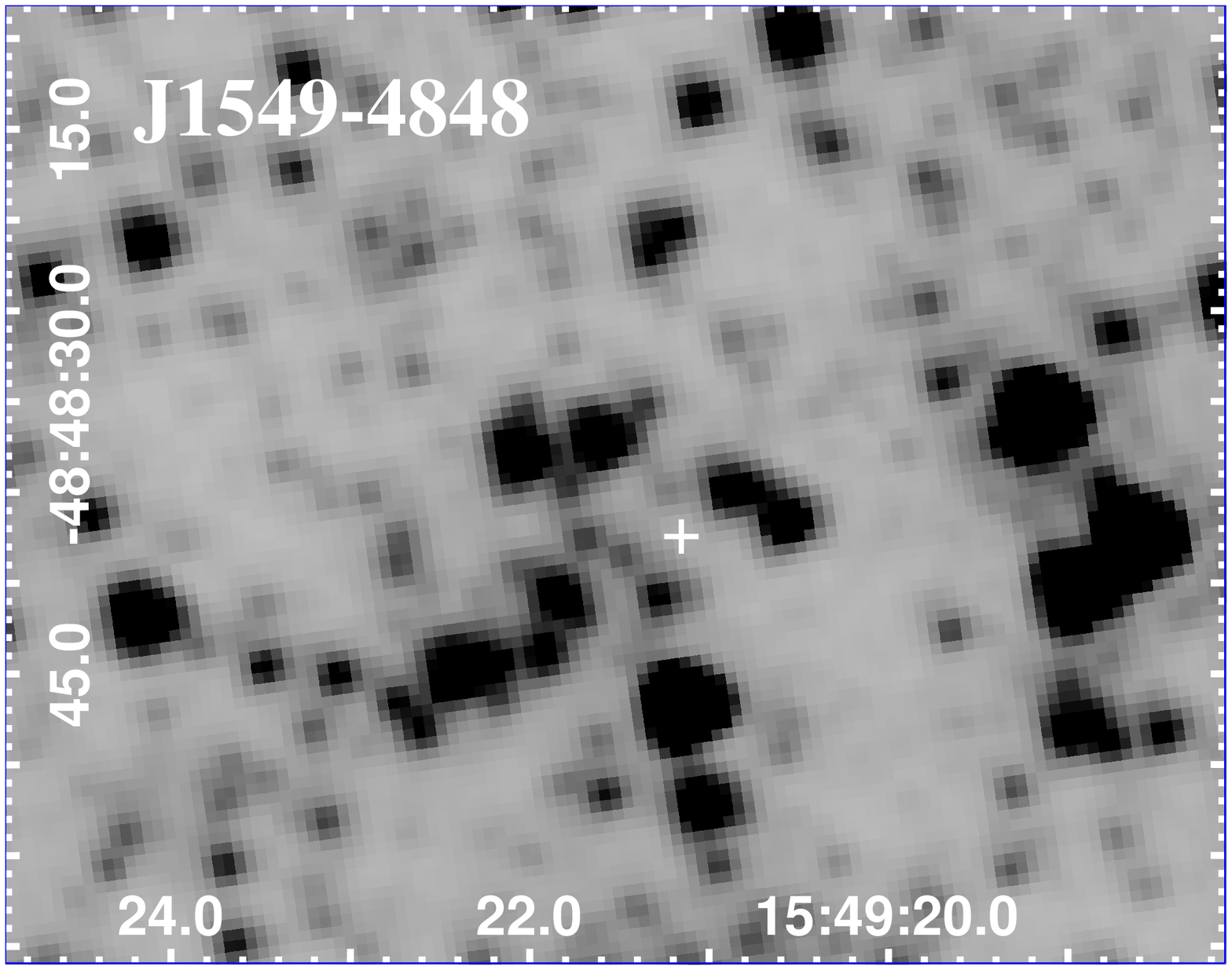} & \\
\end{tabular}
\caption{\textit{Spitzer}/IRAC 4.5~$\mu$m images of the fields of the seven 
pulsar targets. The positions of the pulsars are marked by the crosses.}
\label{fig:tgt}
%%\end{center}
\end{figure*}

For flux calibration, 4-14 relatively bright 2MASS stars detected in 
the fields of the targets were used; bright stars were all saturated in 
our images. The uncertainties are in a range of 0.02--0.06~mag.
For B0950+08, only one star with the 2MASS $K_s$ measurement was detected 
in our image, but the uncertainty is 0.16~mag. Since it was observed
in the same night as J0729$-$1448, the flux calibration results
for the latter were used instead.

\subsection{\textit{Spitzer} Imaging}

The \spz\ observations were conducted in the years of 2007--2008, with the 
observation dates given in Table~\ref{tab:obs}.
The imaging instrument used was the Infrared Array 
Camera (IRAC; \citealt{fha+04}). It operates in four channels 
at  3.6, 4.5, 5.8, and 8.0 $\mu$m.
We observed our targets in Channels 2 (4.5 $\mu$m) and 4 (8.0 $\mu$m).
The detectors at the short and long wavelengths are InSb
and Si:As devices, respectively, with 256$\times$256 pixels and a plate
scale of 1\farcs2 pixel$^{-1}$. The field of view (FOV) 
is 5\farcm2$\times$5\farcm2.
To avoid possible saturation caused by bright stars in a target field,
the frame times were set to either 12 or 30 s, with 10.4 and 26.8 s effective 
exposure time per frame, respectively. The total on-source
time in each observation is given in Table~\ref{tab:obs}.

The raw image data were processed through the IRAC data pipelines 
at the {\em Spitzer} Science Center (SSC). 
In the Basic Calibrated Data (BCD) pipeline, standard imaging data
reductions, such as removal of the electronic bias, dark sky subtraction,
flat-fielding, and linearization, were performed and individual flux-calibrated
BCD frames were produced.  The detailed reduction in the pipelines can be 
found in the IRAC Instrument Handbook\footnote{http://irsa.ipac.caltech.edu/data/SPITZER/docs/irac/iracinstrumenthandbook/}.
Using {\tt MOPEX}, an SSC's package for reducing and analyzing imaging data,
the BCD images were then combined into a final post-BCD (PBCD) mosaic at each 
channel.

%%The sensitivity of the IRAC observations is limited by background
%%sky emission and confusion noise when a field is crowded. 
%%We derived the 3$\sigma$ flux upper limits from the standard
%%deviation of the background sky at the source positions. The upper limits
%%are given in Table~\ref{tab:obs}.
%%As can be seen, \eaxp\  has 
%%a background between medium to high\footnote{See
%%www.spitzer.caltech.edu/obs/bg.html} (i.e., for example at 8 $\mu$m, the sky brightness is between 7.7--18.3 MJy/sr).
%%The backgrounds of the other two AXPs are 
%%much higher than the high background defined by \spitzer, presumably because of their crowded fields.

\subsection{\textit{WISE} Imaging}

Launched on 2009 December 14, \textit{WISE}
mapped the entire sky at 3.4, 4.6, 12, and 22 $\mu$m (called W1, 
W2, W3, and W4 bands, respectively) in 2010 with FWHMs of 
6.1\arcsec, 6.4\arcsec, 6.5\arcsec, and 12.0\arcsec\  in the four bands, 
respectively (see \citealt{wri+10} for details). 
The \textit{WISE} all-sky images and source catalog were released in 2012 March.
We downloaded the \textit{WISE} image data of each source field 
from the Infrared Processing and Analysis Center (IPAC).
The \textit{WISE} observation dates (in year 2010) and
the depths of coverage (in units of pixels) are given in Table~\ref{tab:obs}.
One pixel of the depth of coverage corresponds to 8.8~s on-source integration 
time \citep{wri+10}.

\subsection{X-ray Data Analysis}

There are pointed X-ray observations for six pulsar targets 
(except J1549$-$4848).  Among them, only PSRs J0729$-$1448 and B0950+08
have been detected in X-rays and the results were reported by
\citet{kp09} and \citet{zp04} (see also \citealt{bec+04}), respectively. 
In addition, the X-ray non-detection of PSR J1317$-$5759 has been 
reported by \citet{mcl09}. 

For the rest of the pulsars, there are archival \textit{XMM-Newton}
observations of PSR B0740$-$28 and 
\textit{Swift} observations of PSRs J0940$-$5428 and J1015$-$5719.
PSR J1549$-$4848 was only covered 
by a short exposure in the \textit{ROSAT} All Sky Survey.

We reprocessed the \textit{XMM-Newton} data (ObsID: 0103262401)\footnote{We 
note that the other observation (ObsID: 0103260501) is not usable due to 
severe background flares.} of PSR B0740$-$28 using SAS 13.5.0. 
After removing periods of high
background, we obtained net exposures of 1990\,s, 1630\,s, and 300\,s for
MOS1, MOS2, and PN, respectively. The PN data were not included in this
analysis, because the exposure is too short to be useful. We combined the MOS1
and MOS2 data but found no detection of the pulsar. Using the CIAO task
\texttt{aprates}, which is based on a Bayesian algorithm, we obtained a
3$\sigma$ count rate limit of $7.9\times10^{-3}$\,cts\,s$^{-1}$ in the
0.5--7\,keV energy range. 

For PSRs J0940$-$5428 and J1015$-$5719, we analyzed the \textit{Swift} data taken in
the photon counting mode and processed by the standard pipeline. Three data
sets (ObsIDs: 00036695001, 00036695002, and 00036695003) were used for the
former with a total exposure of 23.7\,ks, and the latter has one 2.8-ks
archival exposure (ObsID: 00031631002). No X-ray counterparts were detected.
The task \texttt{aprates} gives 3$\sigma$ count rate limits of
$5.3\times10^{-4}$\,cts\,s$^{-1}$ and of $4.5\times10^{-3}$\,cts\,s$^{-1}$ for
PSRs J0940$-$5428 and J1015$-$5719, respectively, in the 0.5--7\,keV range.

Finally, the ROSAT survey image shows no detection for PSR J1549$-$4848
in X-rays.  The exposure of the target field is short, only 330\,s.

\section{RESULTS} 
\label{sec:res}

No candidate counterparts to the seven pulsars were detected from 
IR imaging. The positions of the pulsars 
in the \spz/IRAC 4.5 $\mu$m images of the fields are shown
in Figure~\ref{fig:tgt}.
For PSR B0740$-$28, an object is located 0\farcs88 north to the pulsar's 
position, whose uncertainty is only 0\farcs05 determined
from long-term pulsar timing \citep{hob+04}.
Because the object was also detected in our $K_s$ image, which implies
that it is $>$5$\sigma$ away from the pulsar's position, it can
be excluded as a candidate counterpart.
Among the pulsar targets, one IR bow shock was detected
around PSR 1549$-$4848, and studies of the bow shock were 
reported in \citet{wan+13}.

We estimated 3$\sigma$ flux upper limits for our ground-based and \spz\ imaging
by considering that a point source would have been detected if
its peak brightness was 3 times larger than the standard deviation of 
the background counts near a source position.
The flux upper limits, particularly for the IRAC observations, 
are mainly determined by background
sky emission and confusion noise when a field is crowded.
The obtained flux upper limits are given in Table~\ref{tab:obs}.
For \textit{WISE} imaging, images of each source field at the four bands were 
analyzed and bright \textit{WISE} sources were used for flux calibration. 
The \textit{WISE} flux upper
limits for each pulsar were also estimated, which are given in 
Table~\ref{tab:obs}. 
For PSRs J0940$-$5428 and J1015$-$5719, large-scale, nebula-like structures
were detected around the sources from 4.5 $\mu$m to longer wavelength bands
(Figure~\ref{fig:tgt}), and for PSR J1317$-$5759, the background emission 
is high (approximately 10 times higher than that, for example, in the field of
B0950+08) and the field is crowded. These factors resulted in larger flux upper 
limits than that normally obtained.

The unabsorbed X-ray flux upper limits on PSRs B0740$-$28, J09404$-$5428, 
J1015$-$5719, and J1549$-$4848 were also estimated.
For PSR B0740$-$28, we generated the \textit{XMM-Newton} telescope response 
files with SAS and assumed an absorbed power-law spectrum with photon index 
of 1.5. The column density was estimated
from the pulsar dispersion measure (DM) by assuming 10\% 
ionization \citep*{hnk13}. 
An unabsorbed flux limit of $1.2\times10^{-13}$\,erg\,s$^{-1}$~cm$^{-2}$ 
in 0.5--7\,keV was obtained.
With the same procedure, we obtained 
$2.9\times10^{-14}$\,erg\,s$^{-1}$~cm$^{-2}$
and $3.2\times10^{-13}$\,erg\,s$^{-1}$~cm$^{-2}$ for PSRs J0940$-$5428 
and J1015$-$5719, respectively, using the \textit{Swift} telescope response 
files provided by the calibration team.
For PSR 1549$-$4848, we deduced a 3$\sigma$ flux upper limit
of $4.3\times10^{-12}$\,erg\,s$^{-1}$~cm$^{-2}$ (0.1--2.4\,keV) 
using PIMMS from the count rate limit, 
which is not very constraining due to the short exposure.
The unabsorbed X-ray fluxes or flux upper limits for our pulsar
targets are summarized in Table~\ref{tab:tgt}.
%%\begin{figure}
\begin{center}
\includegraphics[scale=0.70]{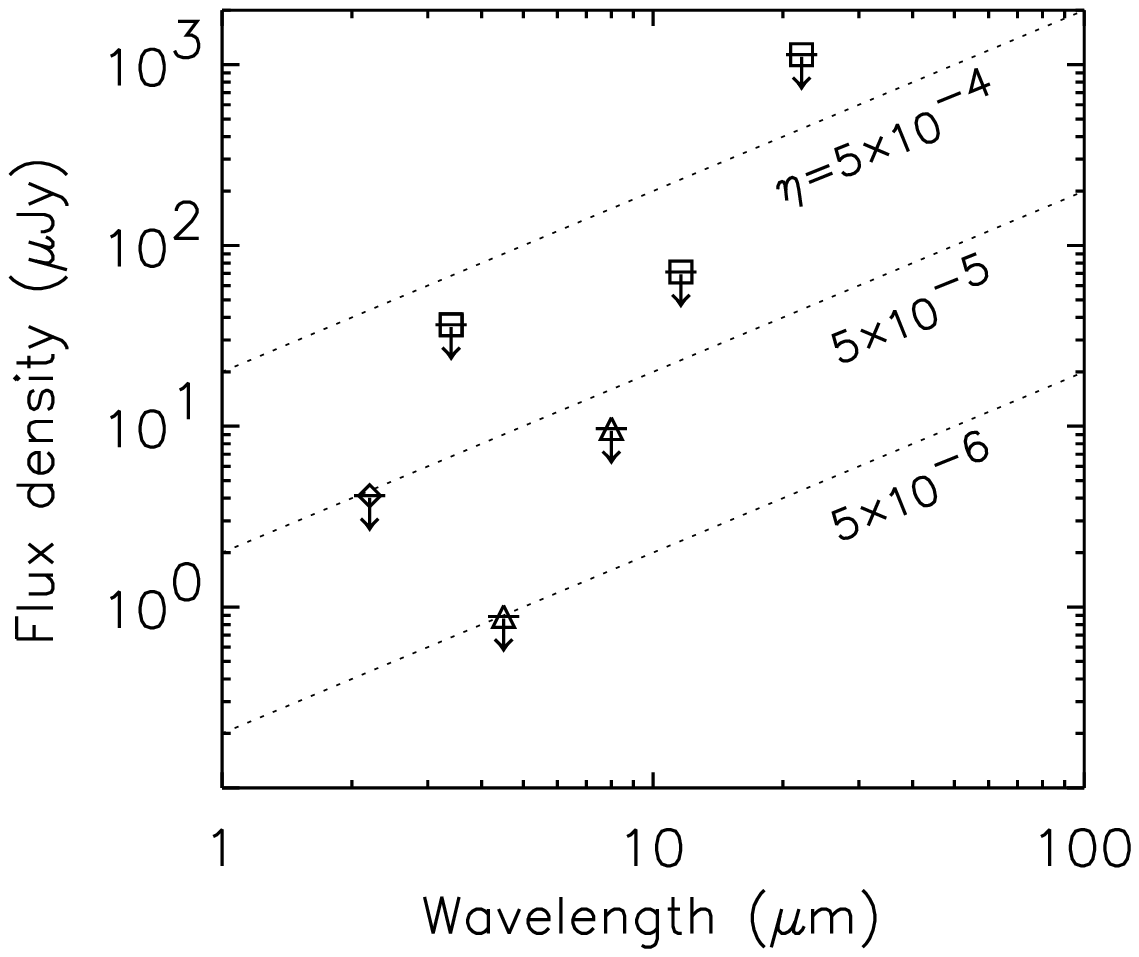}
%%\plotone{f2.eps}
\figcaption{Dereddened ground-based $K_s$ (diamond), \spz\ (triangles), and
\textit{WISE} (squares) flux upper limits of PSR J0729$-$1448. Different 
fraction values of $L_{\rm sd}$ are indicated by dotted lines. 
\label{fig:obs}}
\end{center}
%%\end{figure}

\section{DISCUSSION}
\label{sec:dis}

For the pulsar targets, their $L_{\rm sd}$ and 
$d$ (estimated from their DMs and generally have 30\% uncertainties)
are listed in Table~\ref{tab:tgt}.
Except PSR J1317$-$5759, the pulsars have spin-down
fluxes of 7--90$\times 10^{-11}$ erg s$^{-1}$ cm$^{-2}$. If a debris
disk exists around a pulsar, the energy flux of the pulsar might be 
intercepted and a fraction of it
would be re-radiated by the disk in the IR (e.g., \citealt{ff96}).
Given that the flux range is within an order of magnitude, we use the first
pulsar, PSR J0729$-$1448, in Table~\ref{tab:tgt} as a main example and 
discuss the implications from our results.

The extinction $A_V$ to each pulsar target was estimated using the 
three-dimensional extinction maps of the Galaxy \citep*{dcl03}.
The obtained $A_V$ values are given in Table~\ref{tab:tgt}, which are
generally consistent with those initially estimated from the Galactic disk
dust map (see Section~\ref{subsec:ts}).
In order to obtain the dereddened flux upper limits, based on which 
discussions are made below,
the reddening laws of \citet*{sfd98} for the near-IR $K_s$ band data,
and \citet{ind+05} (wavelengths $\leq$8 $\mu$m) and \citet{wd01}
(wavelengths $>$8 $\mu$m) for \textit{Spitzer} and \textit{WISE} data were used.

In Figure~\ref{fig:obs}, the dereddened flux upper limits for 
PSR J0729$-$1448 are shown. 
Assuming that a fraction $\eta$
of the energy flux from this pulsar would be radiated at the IR bands 
due to the presence of a dust disk
[$\eta(L_{\rm sd}/4\pi d^2)=\nu F_{\nu}$, where
$F_{\nu}$ is the flux density at a band of frequency $\nu$], 
the upper limits have
reached $\eta\sim 5\times 10^{-4}$ at 22~$\mu$m. The $\eta$ value is
nearly as deep as that for the planetary system PSR B1257+12 obtained
from \spz\ MIPS observations \citep{bry+06}. However, as pointed out by 
\citet{bry+06}, the $\eta$ limit is still not sufficiently deep since
most main-sequence stars with debris disks, which include our solar system,
have $\eta < 10^{-5}$. Even if a dust disk exists, emission would not
be strong enough to be detected. 

At 4.5 $\mu$m, our upper limit has reached
$\eta \simeq 5\times 10^{-6}$. This ratio is at least one order of magnitude 
deeper than that of the disk case of the magnetar 4U~0142+61, 
where $(\nu_{4.5\mu}F_{4.5\mu})/F_{\rm X}\sim 10^{-4}$ \citep{wkc07,wkh07}
is used for the comparison
because the main energy flux from 4U~0142+61 (as well as 1E~2259+586) is X-ray 
emission.  In \citet{wck06}, the IR component in the SED of 4U~0142+61 was 
modelled to arise from a passive, X-ray irradiated disk 
(see also \citealt{vrt+90,phn00}), and as the IR--to--X-ray flux ratio
$(\nu_{4.5\mu}F_{4.5\mu})/F_{\rm X}$ might be typical for disks around
X-ray bright neutron stars, the flux ratio has been used for the discussion of 
disk detectability around different young neutron stars 
(e.g., \citealt{wkc07,wkh07}).
%%which is presumably optically 
%%thick at \textit{Spitzer}/IRAC IR channels. A mass upper limit was estimated
%%based on the assumption that optically thin emission would be detected at
%%longer wavelengths with a flux lower than that at the IR channels. 
\citet{cs08} also have proposed that X-ray emission from a neutron star, 
due to thermal radiation from its high-temperature surface and/or non-thermal 
radiation from its magnetosphere, provides heating of a debris disk. 
To check the possibility that a debris disk would be seen mainly due to
X-ray heating, we calculated the $(\nu_{4.5\mu}F_{4.5\mu})/F_{\rm X}$ flux 
ratio upper limits for PSRs J07294$-$1448 and B0950+08, which
are the only two targets detected with X-ray emission. The ratios are
5.9$\times 10^{-2}$ and 3.1$\times 10^{-4}$, 
respectively, suggesting that our observations were not sufficiently
deep. For the pulsar targets without X-ray detection, the X-ray upper 
limits of B0740$-$28, J0940$-$5428, and J1015$-$5719 are below or nearly equal
to 10$^{-3} L_{\rm sd}$, which is approximately the average fraction
of pulsars' spin-down energy emitted at X-rays \citep{bt97,kar+12}. 
The flux limits
and fluxes suggest that at least half of our pulsar targets (including
PSR J0729$-$1448) are unfortunately intrinsically faint at X-ray 
energies.

In this survey of searching for debris disks around neutron stars, we have
focused on relatively young radio pulsars with high spin-down fluxes. While
only seven such pulsars were observed with the most sensitive IR telescope 
due to limited, expensive observing time on a space telescope, the 
results may suggest that it is difficult to detect any debris disks around
radio pulsars, whose existence has been hinted by various pieces of
indirect evidence. Considering the detections of possible
debris disks around two magnetars, another search focusing on
X-ray bright pulsars may be conducted with the future IR telescopes
such as the \textit{James-Webb Space Telescope}. Our survey is also limited to 
relatively hot debris disks, as the stringent sensitivities were provided 
by \spz\ 4.5 and 8.0~$\mu$m imaging. In order to complete this survey at
multiple wavelengths, searches for cold
dust around our pulsar targets should be conducted, given the
great capability now available from 
the Atacama Large Millimeter/submillimeter Array (ALMA) at submillimeter
wavelengths. With its sensitivities at least an order of magnitude 
better than that in the previous searches (e.g., \citealt{pc94,lww04}), 
ALMA observations will help draw a full conclusion about
our searches for debris disks around isolated radio pulsars.

\acknowledgements
This paper includes data gathered with the 6.5 meter Magellan Telescopes 
located at Las Campanas Observatory, Chile.
The work is based in part on observations made with the \textit{Spitzer} 
Space Telescope, which is operated by the Jet Propulsion Laboratory, 
California  Institute of Technology under a contract with NASA.
The publication makes use of data products from the Wide-field Infrared  
Survey Explorer, which is a joint project of the University of California,  
Los Angeles, and the Jet Propulsion Laboratory/California Institute of  
Technology, funded by NASA. 

We thank the anonymous referee for constructive suggestions.
This research was supported by the National Natural Science Foundation of 
China (11373055) and the Strategic Priority Research Program 
``The Emergence of Cosmological Structures" of the Chinese Academy 
of Sciences (Grant No. XDB09000000). Z.W. is a Research Fellow of the 
One-Hundred-Talents project of Chinese Academy of Sciences.
A.L. is supported by NSF AST-1311804 and NASA.

{\it Facilities:} \facility{Magellan (PANIC), \textit{Spitzer} (IRAC)} 
%%{\it Facility:} \facility{Gemini:South}

\bibliographystyle{apj}
%%\bibliography{pdisk}

\begin{thebibliography}{61}
\expandafter\ifx\csname natexlab\endcsname\relax\def\natexlab#1{#1}\fi

\bibitem[{{Becker} \& {Truemper}(1997)}]{bt97}
{Becker}, W., \& {Truemper}, J. 1997, \aap, 326, 682

\bibitem[{{Becker} {et~al.}(2004){Becker}, {Weisskopf}, {Tennant}, {Jessner},
  {Dyks}, {Harding}, \& {Zhang}}]{bec+04}
{Becker}, W., {Weisskopf}, M.~C., {Tennant}, A.~F., {Jessner}, A., {Dyks}, J.,
  {Harding}, A.~K., \& {Zhang}, S.~N. 2004, \apj, 615, 908

\bibitem[{{Blackman} \& {Perna}(2004)}]{bp04}
{Blackman}, E.~G., \& {Perna}, R. 2004, \apjl, 601, L71

\bibitem[{{Bryden} {et~al.}(2006){Bryden}, {Beichman}, {Rieke}, {Stansberry},
  {Stapelfeldt}, {Trilling}, {Turner}, \& {Wolszczan}}]{bry+06}
{Bryden}, G., {Beichman}, C.~A., {Rieke}, G.~H., {Stansberry}, J.~A.,
  {Stapelfeldt}, K.~R., {Trilling}, D.~E., {Turner}, N.~J., \& {Wolszczan}, A.
  2006, \apj, 646, 1038

\bibitem[{{{\c C}ali{\c s}kan} {et~al.}(2013){{\c C}ali{\c s}kan}, {Ertan},
  {Alpar}, {Tr{\"u}mper}, \& {Kylafis}}]{cal+13}
{{\c C}ali{\c s}kan}, {\c S}., {Ertan}, {\"U}., {Alpar}, M.~A., {Tr{\"u}mper},
  J.~E., \& {Kylafis}, N.~D. 2013, \mnras, 431, 1136

\bibitem[{{Chatterjee} {et~al.}(2000){Chatterjee}, {Hernquist}, \&
  {Narayan}}]{chn00}
{Chatterjee}, P., {Hernquist}, L., \& {Narayan}, R. 2000, \apj, 534, 373

\bibitem[{{Chevalier}(1989)}]{che89}
{Chevalier}, R.~A. 1989, \apj, 346, 847

\bibitem[{{Cordes} \& {Shannon}(2008)}]{cs08}
{Cordes}, J.~M., \& {Shannon}, R.~M. 2008, \apj, 682, 1152

\bibitem[{{Draine} \& {Lee}(1984)}]{dl84}
{Draine}, B.~T., \& {Lee}, H.~M. 1984, \apj, 285, 89

\bibitem[{{Drimmel} {et~al.}(2003){Drimmel}, {Cabrera-Lavers}, \&
  {L{\'o}pez-Corredoira}}]{dcl03}
{Drimmel}, R., {Cabrera-Lavers}, A., \& {L{\'o}pez-Corredoira}, M. 2003, \aap,
  409, 205

\bibitem[{{Espinoza} {et~al.}(2011){Espinoza}, {Lyne}, {Kramer}, {Manchester},
  \& {Kaspi}}]{esp+11}
{Espinoza}, C.~M., {Lyne}, A.~G., {Kramer}, M., {Manchester}, R.~N., \&
  {Kaspi}, V.~M. 2011, \apjl, 741, L13

\bibitem[{{Farihi} {et~al.}(2009){Farihi}, {Jura}, \& {Zuckerman}}]{fjz09}
{Farihi}, J., {Jura}, M., \& {Zuckerman}, B. 2009, \apj, 694, 805

\bibitem[{{Fazio} {et~al.}(2004)}]{fha+04}
{Fazio}, G.~G., {et~al.} 2004, \apjs, 154, 10

\bibitem[{{Foster} \& {Fischer}(1996)}]{ff96}
{Foster}, R.~S., \& {Fischer}, J. 1996, \apj, 460, 902

\bibitem[{{Gaensler} \& {Slane}(2006)}]{gs06}
{Gaensler}, B.~M., \& {Slane}, P.~O. 2006, \araa, 44, 17

\bibitem[{{Gotthelf} {et~al.}(2013){Gotthelf}, {Halpern}, \& {Alford}}]{gha13}
{Gotthelf}, E.~V., {Halpern}, J.~P., \& {Alford}, J. 2013, \apj, 765, 58

\bibitem[{{He} {et~al.}(2013){He}, {Ng}, \& {Kaspi}}]{hnk13}
{He}, C., {Ng}, C.-Y., \& {Kaspi}, V.~M. 2013, \apj, 768, 64

\bibitem[{{Hobbs} {et~al.}(2004){Hobbs}, {Lyne}, {Kramer}, {Martin}, \&
  {Jordan}}]{hob+04}
{Hobbs}, G., {Lyne}, A.~G., {Kramer}, M., {Martin}, C.~E., \& {Jordan}, C.
  2004, \mnras, 353, 1311

\bibitem[{{Indebetouw} {et~al.}(2005){Indebetouw}, {Mathis}, {Babler}, {Meade},
  {Watson}, {Whitney}, {Wolff}, {Wolfire}, {Cohen}, {Bania}, {Benjamin},
  {Clemens}, {Dickey}, {Jackson}, {Kobulnicky}, {Marston}, {Mercer},
  {Stauffer}, {Stolovy}, \& {Churchwell}}]{ind+05}
{Indebetouw}, R., {et~al.} 2005, \apj, 619, 931

\bibitem[{{Jones}(2007)}]{jon07}
{Jones}, P.~B. 2007, \mnras, 382, 871

\bibitem[{{Jones}(2008)}]{jon08}
---. 2008, \mnras, 386, 505

\bibitem[{{Jura} {et~al.}(2007){Jura}, {Farihi}, {Zuckerman}, \&
  {Becklin}}]{jur+07}
{Jura}, M., {Farihi}, J., {Zuckerman}, B., \& {Becklin}, E.~E. 2007, \aj, 133,
  1927

\bibitem[{{Kaplan} {et~al.}(2009){Kaplan}, {Chakrabarty}, {Wang}, \&
  {Wachter}}]{kap+09}
{Kaplan}, D.~L., {Chakrabarty}, D., {Wang}, Z., \& {Wachter}, S. 2009, \apj,
  700, 149

\bibitem[{{Kargaltsev} {et~al.}(2012){Kargaltsev}, {Durant}, {Pavlov}, \&
  {Garmire}}]{kar+12}
{Kargaltsev}, O., {Durant}, M., {Pavlov}, G.~G., \& {Garmire}, G. 2012, \apjs,
  201, 37

\bibitem[{{Kargaltsev} \& {Pavlov}(2009)}]{kp09}
{Kargaltsev}, O., \& {Pavlov}, G.~G. 2009, \apj, 702, 433

\bibitem[{{Kaspi} {et~al.}(2006){Kaspi}, {Roberts}, \& {Harding}}]{krh06}
{Kaspi}, V.~M., {Roberts}, M.~S.~E., \& {Harding}, A.~K. 2006, in Compact
  stellar X-ray sources, ed. W.~H.~G. {Lewin} \& M.~{van der Klis}, 279--339

\bibitem[{{Koch-Miramond} {et~al.}(2002){Koch-Miramond}, {Haas}, {Pantin},
  {Podsiadlowski}, {Naylor}, \& {Sauvage}}]{koc+02}
{Koch-Miramond}, L., {Haas}, M., {Pantin}, E., {Podsiadlowski}, P., {Naylor},
  T., \& {Sauvage}, M. 2002, \aap, 387, 233

\bibitem[{{Lazio} \& {Fischer}(2004)}]{lf04}
{Lazio}, T.~J.~W., \& {Fischer}, J. 2004, \aj, 128, 842

\bibitem[{{Lin} {et~al.}(1991){Lin}, {Woosley}, \& {Bodenheimer}}]{lwb91}
{Lin}, D.~N.~C., {Woosley}, S.~E., \& {Bodenheimer}, P.~H. 1991, \nat, 353, 827

\bibitem[{{Livingstone} {et~al.}(2007){Livingstone}, {Kaspi}, {Gavriil},
  {Manchester}, {Gotthelf}, \& {Kuiper}}]{liv+07}
{Livingstone}, M.~A., {Kaspi}, V.~M., {Gavriil}, F.~P., {Manchester}, R.~N.,
  {Gotthelf}, E.~V.~G., \& {Kuiper}, L. 2007, \apss, 308, 317

\bibitem[{{L{\"o}hmer} {et~al.}(2004){L{\"o}hmer}, {Wolszczan}, \&
  {Wielebinski}}]{lww04}
{L{\"o}hmer}, O., {Wolszczan}, A., \& {Wielebinski}, R. 2004, \aap, 425, 763

\bibitem[{{Manchester} {et~al.}(2005){Manchester}, {Hobbs}, {Teoh}, \&
  {Hobbs}}]{man+05}
{Manchester}, R.~N., {Hobbs}, G.~B., {Teoh}, A., \& {Hobbs}, M. 2005, \aj, 129,
  1993

\bibitem[{Martini {et~al.}(2004)Martini, Persson, Murphy, Birk, Shectman,
  Gunnels, \& Koch}]{mpm+04}
Martini, P., Persson, S.~E., Murphy, D.~C., Birk, C., Shectman, S.~A., Gunnels,
  S.~M., \& Koch, E. 2004, \procspie, 5492, 1653, (astro-ph/0406666)

\bibitem[{{McLaughlin}(2009)}]{mcl09}
{McLaughlin}, M. 2009, in Astrophysics and Space Science Library, Vol. 357,
  Astrophysics and Space Science Library, ed. W.~{Becker}, 41

\bibitem[{{McLaughlin} {et~al.}(2006){McLaughlin}, {Lyne}, {Lorimer}, {Kramer},
  {Faulkner}, {Manchester}, {Cordes}, {Camilo}, {Possenti}, {Stairs}, {Hobbs},
  {D'Amico}, {Burgay}, \& {O'Brien}}]{mcl+06}
{McLaughlin}, M.~A., {et~al.} 2006, \nat, 439, 817

\bibitem[{{Menou} {et~al.}(2001){Menou}, {Perna}, \& {Hernquist}}]{mph01}
{Menou}, K., {Perna}, R., \& {Hernquist}, L. 2001, \apj, 559, 1032

\bibitem[{{Michel} \& {Dessler}(1981)}]{md81}
{Michel}, F.~C., \& {Dessler}, A.~J. 1981, \apj, 251, 654

\bibitem[{{Miller} \& {Hamilton}(2001)}]{mh01}
{Miller}, M.~C., \& {Hamilton}, D.~P. 2001, \apj, 550, 863

\bibitem[{{Pavlov} {et~al.}(2004){Pavlov}, {Sanwal}, \& {Teter}}]{pst04}
{Pavlov}, G.~G., {Sanwal}, D., \& {Teter}, M.~A. 2004, in IAU Symposium, Vol.
  218, Young Neutron Stars and Their Environments, ed. F.~{Camilo} \& B.~M.
  {Gaensler}, 239

\bibitem[{{Perna} {et~al.}(2000){Perna}, {Hernquist}, \& {Narayan}}]{phn00}
{Perna}, R., {Hernquist}, L., \& {Narayan}, R. 2000, \apj, 541, 344

\bibitem[{{Phillips} \& {Chandler}(1994)}]{pc94}
{Phillips}, J.~A., \& {Chandler}, C.~J. 1994, \apjl, 420, L83

\bibitem[{{Phinney} \& {Hansen}(1993)}]{ph93}
{Phinney}, E.~S., \& {Hansen}, B.~M.~S. 1993, in Astronomical Society of the
  Pacific Conference Series, Vol.~36, Planets Around Pulsars, ed. J.~A.
  {Phillips}, S.~E. {Thorsett}, \& S.~R. {Kulkarni}, 371--390

\bibitem[{{Qiao} {et~al.}(2003){Qiao}, {Xue}, {Xu}, {Wang}, \& {Xiao}}]{qia+03}
{Qiao}, G.~J., {Xue}, Y.~Q., {Xu}, R.~X., {Wang}, H.~G., \& {Xiao}, B.~W. 2003,
  \aap, 407, L25

\bibitem[{{Reach} {et~al.}(2005){Reach}, {Kuchner}, {von Hippel}, {Burrows},
  {Mullally}, {Kilic}, \& {Winget}}]{rea+05}
{Reach}, W.~T., {Kuchner}, M.~J., {von Hippel}, T., {Burrows}, A., {Mullally},
  F., {Kilic}, M., \& {Winget}, D.~E. 2005, \apjl, 635, L161

\bibitem[{{Reach} {et~al.}(2009){Reach}, {Lisse}, {von Hippel}, \&
  {Mullally}}]{rea+09}
{Reach}, W.~T., {Lisse}, C., {von Hippel}, T., \& {Mullally}, F. 2009, \apj,
  693, 697

\bibitem[{{Schlegel} {et~al.}(1998){Schlegel}, {Finkbeiner}, \&
  {Davis}}]{sfd98}
{Schlegel}, D.~J., {Finkbeiner}, D.~P., \& {Davis}, M. 1998, \apj, 500, 525

\bibitem[{{Shannon} {et~al.}(2013){Shannon}, {Cordes}, {Metcalfe}, {Lazio},
  {Cognard}, {Desvignes}, {Janssen}, {Jessner}, {Kramer}, {Lazaridis},
  {Purver}, {Stappers}, \& {Theureau}}]{sha+13}
{Shannon}, R.~M., {et~al.} 2013, \apj, 766, 5

\bibitem[{{Skrutskie} {et~al.}(2006)}]{2mass}
{Skrutskie}, M.~F., {et~al.} 2006, \aj, 131, 1163

\bibitem[{{Thompson} \& {Duncan}(1996)}]{td96}
{Thompson}, C., \& {Duncan}, R.~C. 1996, \apj, 473, 322

\bibitem[{{Vrtilek} {et~al.}(1990){Vrtilek}, {Raymond}, {Garcia}, {Verbunt},
  {Hasinger}, \& {Kurster}}]{vrt+90}
{Vrtilek}, S.~D., {Raymond}, J.~C., {Garcia}, M.~R., {Verbunt}, F., {Hasinger},
  G., \& {Kurster}, M. 1990, \aap, 235, 162

\bibitem[{{Wang} {et~al.}(2006){Wang}, {Chakrabarty}, \& {Kaplan}}]{wck06}
{Wang}, Z., {Chakrabarty}, D., \& {Kaplan}, D.~L. 2006, \nat, 440, 772

\bibitem[{{Wang} {et~al.}(2007{\natexlab{a}}){Wang}, {Kaplan}, \&
  {Chakrabarty}}]{wkc07}
{Wang}, Z., {Kaplan}, D.~L., \& {Chakrabarty}, D. 2007{\natexlab{a}}, \apj,
  655, 261

\bibitem[{{Wang} {et~al.}(2013){Wang}, {Kaplan}, {Slane}, {Morrell}, \&
  {Kaspi}}]{wan+13}
{Wang}, Z., {Kaplan}, D.~L., {Slane}, P., {Morrell}, N., \& {Kaspi}, V.~M.
  2013, \apj, 769, 122

\bibitem[{{Wang} {et~al.}(2007{\natexlab{b}}){Wang}, {Kaspi}, \&
  {Higdon}}]{wkh07}
{Wang}, Z., {Kaspi}, V.~M., \& {Higdon}, S.~J.~U. 2007{\natexlab{b}}, \apj,
  665, 1292

\bibitem[{{Weingartner} \& {Draine}(2001)}]{wd01}
{Weingartner}, J.~C., \& {Draine}, B.~T. 2001, \apj, 548, 296

\bibitem[{{Wolszczan} \& {Frail}(1992)}]{wf92}
{Wolszczan}, A., \& {Frail}, D.~A. 1992, \nat, 355, 145

\bibitem[{{Woods} \& {Thompson}(2006)}]{wt06}
{Woods}, P.~M., \& {Thompson}, C. 2006, {Soft gamma repeaters and anomalous
  X-ray pulsars: magnetar candidates} (eds.~{Lewin}, W.~H.~G. and {van der
  Klis}, M. (Cambridge: Cambridge Univ.~Press)), 547--586

\bibitem[{{Woosley} \& {Weaver}(1995)}]{ww95}
{Woosley}, S.~E., \& {Weaver}, T.~A. 1995, \apjs, 101, 181

\bibitem[{{Wright} {et~al.}(2010){Wright}, {Eisenhardt}, {Mainzer}, {Ressler},
  {Cutri}, {Jarrett}, {Kirkpatrick}, {Padgett}, {McMillan}, {Skrutskie},
  {Stanford}, {Cohen}, {Walker}, {Mather}, {Leisawitz}, {Gautier}, {McLean},
  {Benford}, {Lonsdale}, {Blain}, {Mendez}, {Irace}, {Duval}, {Liu}, {Royer},
  {Heinrichsen}, {Howard}, {Shannon}, {Kendall}, {Walsh}, {Larsen}, {Cardon},
  {Schick}, {Schwalm}, {Abid}, {Fabinsky}, {Naes}, \& {Tsai}}]{wri+10}
{Wright}, E.~L., {et~al.} 2010, \aj, 140, 1868

\bibitem[{{Xu} \& {Jura}(2012)}]{xj12}
{Xu}, S., \& {Jura}, M. 2012, \apj, 745, 88

\bibitem[{{Zavlin} \& {Pavlov}(2004)}]{zp04}
{Zavlin}, V.~E., \& {Pavlov}, G.~G. 2004, \apj, 616, 452

\end{thebibliography}

\begin{deluxetable}{l c c c c c c c c }
%%\rotate
\tablecolumns{9}
\tablecaption{Properties and derived fluxes or flux upper limits for the seven 
radio pulsar targets}
\tablewidth{0pt}
\tablehead{
\colhead{Pulsar} & \colhead{Age} &  \colhead{$d$} & \colhead{$Gb$} & 
\colhead{$L_{\rm sd}$/10$^{35}$} & \colhead{$F_{\rm X}$\tablenotemark{a}$/10^{-14}$} & \colhead{$A_V$} & \colhead{$F_{4.5\mu}$\tablenotemark{b}} & \colhead{$(\nu_{4.5\mu}F_{4.5\mu})/F_{\rm X}$} \\
\colhead{} & \colhead{(kyr)} & \colhead{(kpc)} &
\colhead{(deg)} & \colhead{(erg s$^{-1}$)} & \colhead{(erg s$^{-1}$ cm$^{-2}$)} & \colhead{(mag)} & ($\mu$Jy) & \colhead{} 
}
\startdata
J0729$-$1448 & 35 & 4.4 & $+$1.4 & 2.8 & 1.0 & 1.1 & $<$0.87 & $< 5.9\times 10^{-2}$ \\
B0740$-$28  & 160 & 1.9 & $-$2.4 & 1.4  & $<$12 & 0.93 & $<$0.061 & \nodata  \\
J0940$-$5428 & 40 & 4.3 & $-$1.3 & 19 & $<$2.9 & 4.1 & $<$1.6 &\nodata    \\
B0950$+$08 &  10$^4$ & 0.26 & $+$43.7 & 0.006 & 12 & 0.088 & $<$0.056 & $< 3.1\times 10^{-4}$ \\
J1015$-$5719 &  40 & 4.9 & $-$0.6 & 8.3 & $<$32 & 5.3 & $<$0.79 & \nodata \\
J1317$-$5759 & 10$^3$ & 5.6 & $+$4.7 & 0.002 & $<$74 & 2.2 & $<$16 & \nodata \\
J1549$-$4848 &  300 & 1.5 & $+$4.3 & 0.2 & $<$430 & 1.6 & $<$0.41 & \nodata \\
\enddata
\tablenotetext{a}{X-ray flux is unabsorbed. The energy range for 
J0729$-$1448 is 0.5--8 keV, for B0740$-$28, J0940$-$5428, and J1015$-$5719 
is 0.5--7 keV, for B0950+08 is 0.2--10 keV, for J1317$-$5759 is 0.3--5 keV,
and for J1549$-$4848 is 0.1--2.4 keV.}
\tablenotetext{b}{Dereddened flux upper limits at the \textit{Spitzer} IRAC channel 2.}
\label{tab:tgt}
\end{deluxetable}

\begin{deluxetable}{l c c c c c c c}
%%\rotate
\tablecolumns{8}
\tablecaption{Observations of seven radio pulsar targets}
\tablewidth{0pt}
\tablehead{\colhead{} & \colhead{J0729$-$1448} & \colhead{B0740$-$28} & 
\colhead{J0940$-$5428} & \colhead{B0950$+$08} & \colhead{J1015$-$5719} & 
\colhead{J1317$-$5759} & \colhead{J1549$-$4848} }
\startdata
\sidehead{Magellan/PANIC}
Obs. date & 2006/05/15 & 2005/10/22 & 2006/05/16 & 2006/05/15 & 2006/05/16 &
2006/05/16 & 2006/05/17\\
Exposure (min) & 22.5 & 13.5 & 22.5 & 30.0 & 22.5 & 30.0 & 22.5\\
FWHM (arcsec) & 0.56 & 0.43 & 0.40 & 0.46 & 0.39 & 0.49 & 0.40\\
$K_s$ flux limit (mag) & 20.7 & 21.2 & 20.7 & 20.7 & 20.4 & 21.0 & 21.2 \\
\tableline
\sidehead{Spitzer/IRAC}
Obs. date &  2007/11/14 &  2008/01/01 &  2008/04/08 &  2007/12/27 & 2008/04/08 & 2007/09/12 &  2007/09/12 \\
Exposure (min) & 17.9 & 17.9 &  17.3 &  17.9 &  17.9 & 17.3 & 17.9 \\
Ch2 flux limit ($\mu$Jy) & 0.82 & 0.058 & 1.3 & 0.056 & 0.61 & 14 & 0.38 \\
Ch4 flux limit ($\mu$Jy) & 9.1 & 6.9 & 89 & 6.4  & 140 & 15 & 1.4 \\
\tableline
\sidehead{WISE}
Obs. date & 2010/04/10 & 2010/04/16 & 2010/06/14 & 2010/05/11 & 2010/01/07 &
2010/01/31 & 2010/02/22\\
  & \nodata & \nodata & \nodata & \nodata & 2010/07/05 & 2010/08/04 & \nodata \\
Depth of coverage (pixel) & 14 & 14 & 32 & 12 & 34 & 34 & 14 \\
W1 flux limit ($\mu$Jy) & 33 & 61 & 160 & 170 & 140 & 94 & 85 \\
W2 flux limit ($\mu$Jy) & 20 & 34 & 87 &  37 &  110 & 27 & 56 \\
W3 flux limit ($\mu$Jy) & 67 & 92 & 250 & 89 & 880 & 72 & 140 \\
W4 flux limit ($\mu$Jy) & 1100 & 1100 & 5400 & 1300 & 5100 & 750 & 790\\
\enddata
\label{tab:obs}
\end{deluxetable}

\clearpage

\end{document}